\documentclass[draftcls,a4paper,onecolumn]{IEEEtran}
\usepackage{cite}
\ifCLASSINFOpdf
  \usepackage[pdftex]{graphicx}
\else
  \usepackage[dvips]{graphicx}
\fi
\usepackage[cmex10]{amsmath}
\usepackage{array}
\usepackage{theorem}
\usepackage[caption=false,font=footnotesize]{subfig}
\usepackage{fixltx2e}
\usepackage{url}
\usepackage{acronym}
\usepackage{fancyhdr}

\makeatletter

{\theorembodyfont{\rmfamily}
   }
{\theorembodyfont{\rmfamily}
   }
{\theorembodyfont{\rmfamily}
   }
{\theorembodyfont{\rmfamily}
   }

\acrodef{LDPC}{Low-density parity-check}
\acrodef{MDPC}{moderate-density parity-check}
\acrodef{QC}{quasi-cyclic}
\acrodef{QC-LDPC}{quasi-cyclic low-density parity-check}
\acrodef{QC-MDPC}{quasi-cyclic moderate-density parity-check}
\acrodef{RSA}{Rivest, Shamir, Adleman}
\acrodef{BF}{bit flipping}
\acrodef{SPA}{sum product algorithm}
\acrodef{RDF}{random difference families}
\acrodef{ISDA}{information set decoding attacks}
\acrodef{DCA}{dual code attacks}
\acrodef{WF}{work factor}
\acrodef{BER}{bit error rate}
\acrodef{CER}{codeword error rate}

\makeatother
\begin{document}

\pagestyle{fancy}
\fancyhf{} % clear all header and footer fields
\fancyhead[L]{\small\textit{To be presented at IEEE ICC 2013 - Workshop on Information Security over Noisy and Lossy Communication Systems \\ Copyright transferred to IEEE}}

\title{Optimization of the parity-check matrix density in QC-LDPC code-based McEliece cryptosystems
\thanks{This work was supported in part by the MIUR project ``ESCAPADE''
(Grant RBFR105NLC) under the ``FIRB -– Futuro in Ricerca 2010'' funding program.}
}

\author{\IEEEauthorblockN{Marco Baldi, Marco Bianchi, Franco Chiaraluce,\\}
\IEEEauthorblockA{DII, Universit\`a Politecnica delle Marche,\\
Ancona, Italy\\
Email: \{m.baldi, m.bianchi, f.chiaraluce\}@univpm.it}}

\maketitle

\thispagestyle{fancy}

\begin{abstract}
\ac{LDPC} codes are one of the most promising families of codes to replace the Goppa codes
originally used in the McEliece cryptosystem. In fact, it has been shown that by using \ac{QC-LDPC} codes in this
system, drastic reductions in the public key size can be achieved, while maintaining fixed security levels.
Recently, some proposals have appeared in the literature using codes with denser parity-check matrices, named \ac{MDPC} codes.
However, the density of the parity-check matrices to be used in \ac{QC-LDPC} code-based variants of the McEliece cryptosystem
has never been optimized.
This paper aims at filling such gap, by proposing a procedure for selecting the density of the private
parity-check matrix, based on the security level and the decryption complexity.
We provide some examples of the system parameters obtained through the proposed technique.
\end{abstract}

\section{Introduction}
\label{sec:Intro}

The perspective of introducing quantum computers has driven a renewed interest towards public-key encryption
schemes which are alternative to widespread solutions, like the \ac{RSA} system, based on the integer factorization problem.
The latter, in fact, would be solved in polynomial time through quantum computers, and hence would no longer
represent a hard problem after their advent.

The McEliece and Niederreiter cryptosystems \cite{McEliece1978, Niederreiter1986}, which exploit the hardness of the decoding
problem to implement public-key cryptography, are among the most interesting alternatives to \ac{RSA}.
Secure instances of these systems are based on Goppa codes and, despite some revision of their parameters
due to optimized cryptanalysis and increased computational power \cite{Bernstein2008},
they have never been seriously endangered by cryptanalysis.
However, using Goppa codes has the major drawback of requiring large public keys, whose size increases
quadratically in the security level.
Several attempts to replace Goppa codes have been made during years, but only a few have resisted cryptanalysis.
Among them, variants based on \ac{QC-LDPC} codes are very promising, since they achieve very
small keys, with size increasing linearly in the security level.
These variants are unbroken up to now, though some refinements have been necessary since their first proposal.

\ac{LDPC} codes are state-of-the-art iteratively decoded codes, first introduced by Gallager \cite{Gallager},
then rediscovered \cite{MacKay99} %, Richardson2001} 
and now used in many contexts \cite{Paolini2009}.
Recently, \ac{LDPC} codes have also been introduced in several security-related contexts, like physical layer security
\cite{Baldi2010, Baldi2011b, Baldi2012a} and key agreement over wireless channels \cite{Renna2013}.
\ac{LDPC} codes were initially thought to be insecure in the McEliece cryptosystem \cite{Monico2000},
and very large codes were required to avoid attacks \cite{Baldi2007ICC}.
This scenario has changed when it has been shown that the permutation matrix used to obtain the public key 
from the private key could be replaced with a more general matrix \cite{Baldi2007ISIT}.
Despite some adjustments have been necessary after the first proposal, these matrices have allowed to design 
secure and efficient instances of the system based on \ac{QC-LDPC} codes \cite{Baldi2008, Baldi2012}.

Recently, it has been shown that the use of permutation matrices, like in the original McEliece cryptosystem, 
can be restored by using codes with increased parity-check matrix density, named \ac{MDPC} codes \cite{Misoczki2012, Biasi2012}.
\ac{MDPC} codes also exhibit performance which does not degrade significantly when
there are short cycles in their associated Tanner graph.
This allows for a completely random code design, which has permitted to obtain a security reduction to the
hard problem of decoding a generic linear code \cite{Misoczki2012}.

In this paper, we compare \ac{LDPC} and \ac{MDPC} code-based McEliece proposals and provide a procedure to optimize the density
of the parity-check matrices of the private code, in such a way as to reach a fixed security level
and, at the same time, keep complexity to the minimum.
The paper is organized as follows:
in Section \ref{sec:ErrCorr}, we assess the error correction performance of the codes of interest, and its dependence on the parity-check matrix density;
in Section \ref{sec:SecLevel}, we estimate the security level of the system by considering the most dangerous structural and local attacks;
in Section \ref{sec:Optimization}, we show how to optimize the private parity-check matrix density by taking into account complexity;
in Section \ref{sec:Examples} we provide some system design examples through the proposed procedure and, finally,
in Section \ref{sec:Conclusion} we draw some conclusive remarks.

\section{Error correction performance}
\label{sec:ErrCorr}

\ac{QC-LDPC} and \ac{QC-MDPC} code-based variants of the McEliece cryptosystem use codes with length $n = n_0 \cdot p$, dimension 
$k = k_0 \cdot p$ and redundancy $r = p$, where $n_0$ is a small integer (\textit{e.g.}, $n_0 = 2,3,4$), $k_0 = n_0-1$, and $p$
is a large integer (on the order of some thousands or more).
The code rate is therefore $\frac{n_0-1}{n_0}$. Since adopting a rather high code rate is important to reduce the encryption
overhead on the cleartext, in this work we focus on the choice $n_0=4$, such that the size of a 
cleartext is $0.75$ times that of the corresponding ciphertext.

The private key contains a \ac{QC} parity-check matrix having the following form \cite{Baldi2011a, Baldi2012}:
\begin{equation}
\mathbf{H} = \left[ \mathbf{H}_{0} | \mathbf{H}_{1} | \ldots |\mathbf{H}_{n_0-1} \right],
\label{eq:HCircRow}
\end{equation}
where each $\mathbf{H}_{i}$ is a circulant matrix with row and column weight $d_v$.
It follows that the row weight of $\mathbf{H}$ is $d_c = n_0 d_v \ll n$.
So, the code defined by $\mathbf{H}$ is an \ac{LDPC} code or \ac{MDPC} code,
according to the definition in \cite{Misoczki2012}.
Actually, the border between \ac{LDPC} and \ac{MDPC} codes is not tidy: \ac{MDPC} codes
are \ac{LDPC} codes too, but their parity-check matrix density is 
not optimal, in regard to the error rate performance.
%higher than that required to achieve the best performance, and hence their performance is suboptimal.

The private key also contains two other matrices: a $k \times k$ non singular scrambling matrix $\mathbf{S}$ 
and an $n \times n$ non singular transformation matrix $\mathbf{Q}$ having average row and column weight $m$.
For the sake of simplicity, $m$ was always chosen as an integer in previous proposals \cite{Baldi2008, Baldi2012},
and $\mathbf{Q}$ was a regular matrix.
However, $\mathbf{Q}$ can also be slightly irregular, in such a way that $m$ can be rational.
%, and the rows and columns of $\mathbf{Q}$ can have two different weights: $\left\lfloor m \right\rfloor$ and $\left\lceil m \right\rceil$.
This provides a further degree of freedom in the design of the system parameters,
which will be exploited in this paper.

Let $\mathbf{G}$ be the private code generator matrix,
the public key is obtained as $\mathbf{G}' = \mathbf{S}^{-1} \cdot \mathbf{G} \cdot{\mathbf{Q}^{-1}}$
for the McEliece cryptosystem and as $\mathbf{H'} = \mathbf{S^{-1} \cdot H \cdot Q}^T$ for the Niederreiter version \cite{Baldi2011}.
In order to preserve the \ac{QC} nature of the public keys, the matrices $\mathbf{S}$ and $\mathbf{Q}$
are also chosen to be \ac{QC}, that is, formed by $k_0 \times k_0$ and $n_0 \times n_0$ circulant blocks, respectively.
This way, and by using a suitable CCA2 secure conversion of the system \cite{Bernstein2008}, which allows using public keys
in systematic form, the public key size becomes
equal to $k_0 \cdot (n_0 - k_0) \cdot p = (n_0 - 1) \cdot p$ bits, which is very small compared to Goppa code-based instances.
On the other hand, the use of $\mathbf{Q}$ in \ac{QC} form limits the resolution of $m$, which cannot vary by less than $1/n_0^2$, 
but this is not an important limitation in the present context.
When using \ac{MDPC} codes, the matrix $\mathbf{Q}$ reduces to a permutation matrix $\mathbf{P}$ (i.e., $m=1$).
In this case, by using a CCA2 secure conversion of the system, 
$\mathbf{S}$ and $\mathbf{P}$ can be eliminated \cite{Misoczki2012}, since the public generator matrix can be in systematic
form and $\mathbf{G}$ can be directly used as the public key.
In fact, differently from Goppa codes, when using \ac{MDPC} codes, exposing $\mathbf{G}$ does not allow an attacker
to perform efficient decoding.

Though the public matrices are dense, the public code admits a valid parity-check matrix in the form $\mathbf{H}' = \mathbf{H} \cdot \mathbf{Q}^T$,
which, due to the sparse nature of both $\mathbf{H}$ and $\mathbf{Q}$, has column and row weight approximately equal to
$d_v' = m d_v$ and $d_c' = m d_c$, respectively.
The matrix $\mathbf{Q}$ has also effect on the intentional error vectors used for encryption, since if Alice adds $t$ intentional
errors for encrypting a message, then Bob must be able to correct up to $t' = m t$ errors to decrypt it \cite{Baldi2012}.

\begin{figure}[tb]
\begin{centering}
\includegraphics[keepaspectratio, width=80mm]{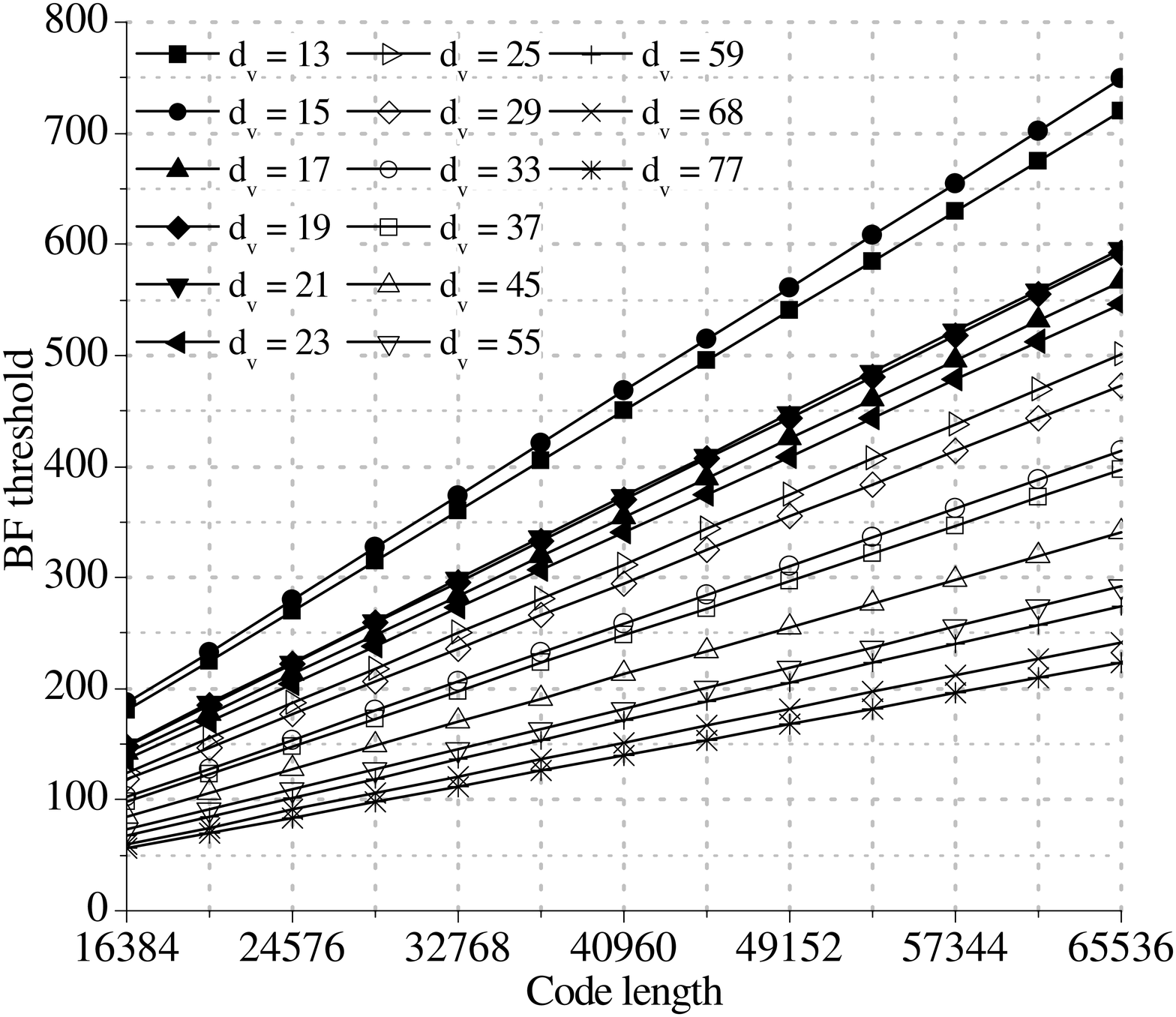}
\caption{\ac{BF} decoding threshold as a function of the code length for $n_0=4$ and several parity-check matrix column weights ($d_v$).}
\label{fig:BFthresholds}
\par\end{centering}
\end{figure}

Concerning the error correction performance of the private code, though for \ac{LDPC} codes its evaluation without simulations is in general
a hard task, we can get a reasonable estimate by computing the \ac{BF} decoding threshold \cite{Baldi2012}.
We have computed this threshold, for $n_0=4$, by considering a fixed and optimized decision threshold for the \ac{BF} decoder,
and letting $p$ vary between $2^{12}$ and $2^{14}$.
Since we are interested in studying the dependence of the \ac{BF} threshold on the parity-check matrix density,
we computed such a threshold for different column weights ($d_v$) ranging between $13$ and $77$.
The results obtained are reported in Fig. \ref{fig:BFthresholds}. We observe that the decoding threshold, so estimated,
increases linearly in the code length, and generally decreases for increasing parity-check matrix densities, though with some local oscillations.

\begin{figure}[tb]
\begin{centering}
\includegraphics[keepaspectratio, width=80mm]{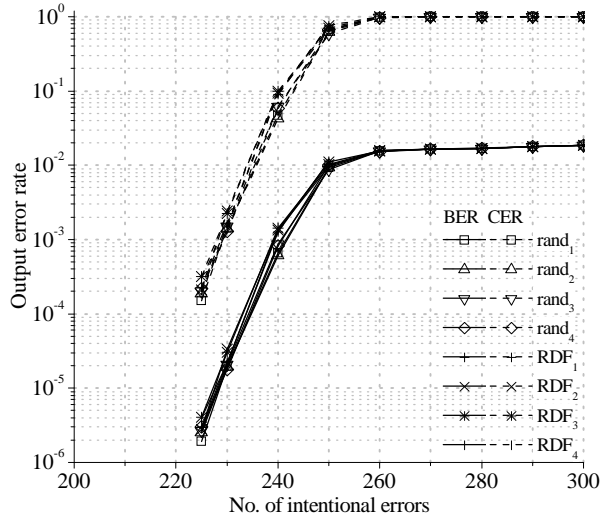}
\caption{Simulated \ac{SPA} decoding performance (\ac{BER} and \ac{CER}) for completely random and \ac{RDF}-based codes with $n_0=4$, $p=4096$ and $d_v=13$.}
\label{fig:Simulations}
\par\end{centering}
\end{figure}

Actually, the \ac{BF} threshold represents the waterfall threshold when using \ac{BF} decoding on an infinite-length code without cycles in the Tanner graph, and hence it does not correspond
to sufficiently low error rates when such a decoding algorithm is used on finite-length codes.
However, several variations and improvements of the \ac{BF} algorithm have been proposed for decoding \ac{LDPC} codes,
and they actually provide very low, and even negligible, residual error rates when the number of errors equals, or slightly overcomes, the \ac{BF} threshold \cite{Baldi2012}.
Even better performance can be achieved by using \ac{LDPC} decoding algorithms based on soft decision, like the \ac{SPA}.
%\cite{Hagenauer1996}.
Thus, for these codes, we can actually use the \ac{BF} threshold as a measure of the number of errors that can be corrected with very high probability.
An example in this sense is provided in Fig. \ref{fig:Simulations}, where the error correcting performance achieved by eight
\ac{QC-LDPC} codes with $n_0=4$, $p=4096$ and $d_v=13$ through \ac{SPA} decoding is reported.
The residual \ac{BER} and \ac{CER} after decoding have been assessed through simulation.
According to Fig. \ref{fig:BFthresholds}, the \ac{BF} threshold for these codes is $181$ errors, and Fig. \ref{fig:Simulations}
confirms that it provides a conservative estimate of the number of correctable errors.
%Actually, their error correction capability turns out to be significantly greater than that estimated through the \ac{BF} threshold,
%as also observed in \cite{Misoczki2012}. Hence, we can consider at least a $10\%$ increase of the \ac{BF} threshold
%for obtaining more realistic evaluations of the actual error correction capability of these codes.

The same conclusion does not seem to be valid for \ac{MDPC} codes, especially for high $n_0$ values.
As an example, we have considered a code 
%suggested in \cite{Misoczki2012} for achieving $128$-bit security 
with $n_0=4$, $n=25088$ and $d_v=85$.
Its \ac{BF} threshold is at $77$ errors; %and the authors suggest to use $68$ errors to achieve a sufficiently low residual error rate. 
however, we have verified through simulations that, with $68$ intentional errors, the \ac{SPA}
achieves a residual \ac{CER} of about $4 \cdot 10^{-3}$.
This result can be improved by resorting to \ac{BF} decoding. In fact, for \ac{MDPC} codes, which have many short cycles
in their Tanner graphs, using soft information may result in worse performance than using good hard-decision
decoding algorithms. For example, the \ac{BF} decoder with variable and optimized decision thresholds is
able to reach a residual \ac{CER} of about $1.5 \cdot 10^{-5}$.
However, these residual error rates confirm that, for \ac{MDPC} codes, the \ac{BF} threshold
may overestimate the number of correctable errors.
%are rather high for the use in the McEliece cryptosystem, hence in Section \ref{sec:Examples} we propose more conservative parameter choices.

From Fig. \ref{fig:Simulations} we also get another important information.
The first four codes considered (denoted by rand$_i, i = 1,\ldots,4$) were designed completely at random, that is, by randomly choosing the positions
of the $13$ ones in the first row of each circulant block.
The second four codes considered (denoted by RDF$_i, i = 1,\ldots,4$) were instead designed by using \ac{RDF} \cite{Baldi2007ISIT}.

From the figure we observe that no significant difference appears between the two sets of curves.
These codes have the lowest parity-check matrix density among those considered, that is, $d_v=13$.
A similar behavior was observed in \cite{Misoczki2012} for \ac{MDPC} codes with $d_v$ on the order of $45$ or more.
This suggests that, for the parity-check matrix densities that are of interest for this kind of applications,
there is no substantial difference between completely random and constrained random code designs.
A difference would instead appear for sparser matrices, like those of interest for application of \ac{LDPC} codes to transmissions
(that is, with $d_v$ on the order of some units), for which short cycles in the Tanner
graph deteriorate the code minimum distance.
Hence, it is reasonable to conclude that a completely random code design can be used in this context, independently of
the parity-check matrix density of the private code. Therefore, the security reduction provided in \cite{Misoczki2012} 
also applies to \ac{LDPC} code-based variants of the McEliece cryptosystem, similarly to those using \ac{MDPC} codes.

\section{Security level}
\label{sec:SecLevel}

%The security level of \ac{QC-LDPC} code-based variants of the McEliece cryptosystem can be assessed
%by considering 
The most dangerous attacks against the considered systems are \ac{DCA} and \ac{ISDA} \cite{Baldi2012}.
In order to estimate the \ac{WF} of these attacks, we consider the algorithm proposed in
\cite{Peters2010} to search for low weight codewords in a random linear code.
Actually, some advances have recently appeared in the literature 
concerning decoding of binary random linear codes \cite{May2011, Becker2012}.
However, these works are more focused on asymptotic evaluations rather than on actual operation counts,
which are needed for our \ac{WF} estimations.
Also ``ball collision decoding'', proposed in \cite{Bernstein2011}, achieves important \ac{WF}
reductions asymptotically, but these reductions are negligible for the considered code lengths and security levels.

\begin{figure}[tb]
\begin{centering}
\includegraphics[keepaspectratio, width=80mm]{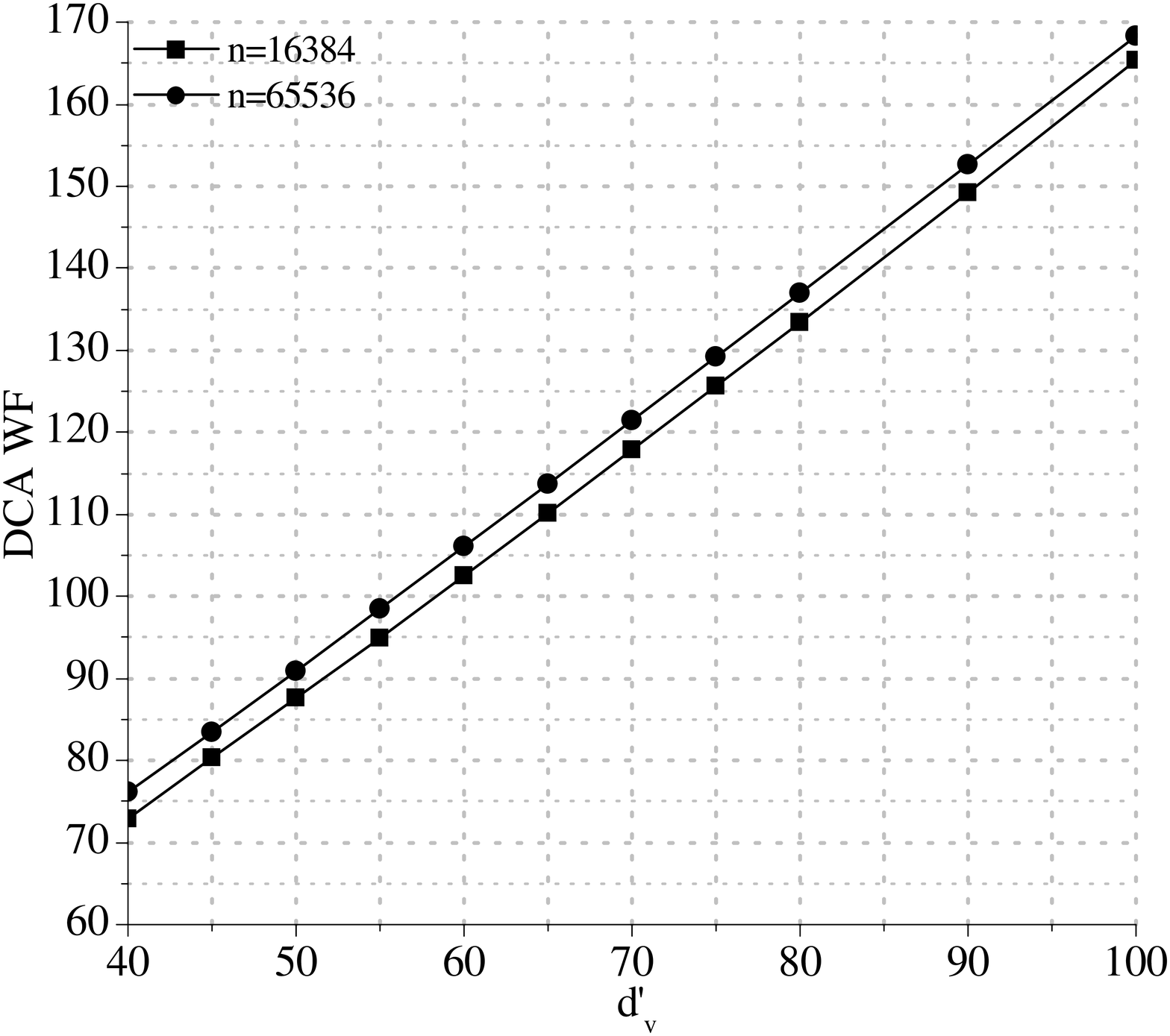}
\caption{\ac{DCA} \ac{WF} ($\log_2$) as a function of the public parity-check matrix column weight ($d_v'$), for $n_0=4$ and $p=4096,16384$.}
\label{fig:DCA}
\par\end{centering}
\end{figure}

\ac{DCA} aim at obtaining the private key from the public key by searching for low weight codewords
in the dual of the public code. This way, an attacker could find the rows of $\mathbf{H'}$, and then
use $\mathbf{H'}$, which is sparse, to decode the public code through \ac{LDPC} decoding algorithms.
The row weight of $\mathbf{H'}$ is $d_c' = n_0 d_v'$ and the corresponding multiplicity is
$r = p$. Figure \ref{fig:DCA} reports the values of the \ac{WF} of \ac{DCA}, as functions
of $d_v'$, for the shortest and the longest code lengths here considered.
We observe that, for a fixed $d_v'$, the two curves differ by less than $2^4$, hence \ac{DCA} exhibit
a weak dependence on $n$.

\begin{figure}[tb]
\begin{centering}
\includegraphics[keepaspectratio, width=80mm]{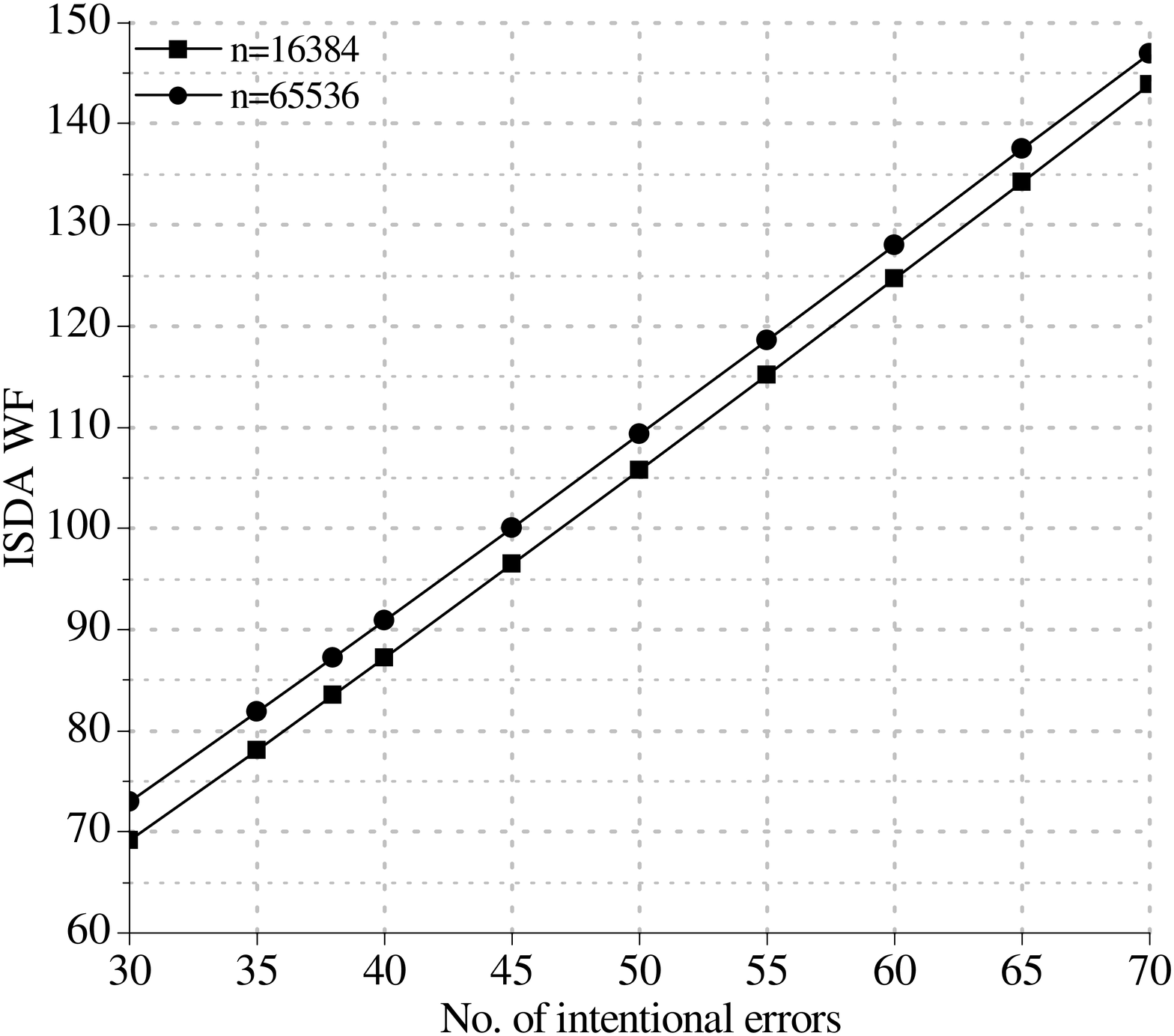}
\caption{\ac{ISDA} \ac{WF} ($\log_2$) as a function of the number of intentional errors, for $n_0=4$ and $p=4096,16384$.}
\label{fig:ISDA}
\par\end{centering}
\end{figure}

\ac{ISDA} instead aim at finding the error vector $\mathbf{e}$ affecting an intercepted ciphertext. 
This can be done by searching for the minimum weight codewords of the extended code generated by
$\mathbf{G}''=\left[\begin{array}{c}\mathbf{G}'\\\mathbf{x}\end{array}\right]$.
This task is facilitated by the \ac{QC} nature of the codes we consider, since each block-wise 
cyclically shifted version of an intercepted ciphertext is another valid ciphertext.
Hence, $\mathbf{G}''$ can be further extended by adding block-wise shifted versions of the
intercepted ciphertext, and the attacker can search for one among as many shifted 
versions of the error vector.
We have considered the optimum number of shifted ciphertexts that can be used by an attacker,
and computed the \ac{WF} of \ac{ISDA} according to the above procedure.
The results obtained are reported in Fig. \ref{fig:ISDA}, as functions of the number of intentional
errors, for the smallest and the largest code lengths here considered.
Also in this case, we observe that the \ac{WF} of the attack has a weak dependence on the code length.

From Fig. \ref{fig:ISDA} we also observe that the \ac{ISDA} \ac{WF} (in $\log_2$) increases
linearly in the number of intentional errors, and we know from Fig. \ref{fig:BFthresholds}
that the decoding threshold increases linearly in the code length.
Hence, provided that $d_v'$ is chosen in such a way that \ac{DCA} have \ac{WF} equal to or
higher than \ac{ISDA}, the security level of the system increases linearly in the code length,
which is a desirable feature for any cryptosystem.

\section{Density optimization}
\label{sec:Optimization}

%In this section we aim at optimizing the density of the private code parity-check matrix
%based on security and complexity considerations.

Some features of the McEliece cryptosystem variants we study are not affected by the private parity-check matrix density.
One of them is the key size. In fact, the public key is always a dense matrix and, hence,
its size does not change between \ac{LDPC} and \ac{MDPC} code-based variants.
The public key size can be reduced to the minimum by using $n_0 = 2$, as in \cite{Misoczki2012}, but this reduces
the code rate to $1/2$, which is less than in the original McEliece cryptosystem and its most recent variants.
We instead consider $n_0 = 4$, which gives slightly larger keys, but also a more sensible code rate.
In fact, due to the \ac{QC} nature of the public matrices, the public key size remains very small,
and increases linearly in the code length, that is, for the considered cryptosystem, in the security level.
Some examples of key size can be found in \cite{Baldi2008, Baldi2012, Misoczki2012}, both
for classical cryptosystem versions and CCA2 secure conversions.

Also the encryption complexity is not affected by the private matrix density, since encryption
is performed through the dense public matrix.
Concerning decryption, the following steps must be performed to decrypt a ciphertext \cite{Baldi2012}:
\begin{enumerate}
\item multiplication of the ciphertext by $\mathbf{Q}$;
\item \ac{LDPC} decoding;
\item multiplication of the decoded information word by $\mathbf{S}$.
\end{enumerate}
The last step is not affected by the private parity-check matrix density, while the complexity of the
first two steps depends on it.
More specifically, the matrix $\mathbf{Q}$ is sparse, hence the cost of step i) is proportional
to its average column weight ($m$). Since, once having fixed $d_v'$ according to the desired security
level against \ac{DCA}, $m$ equals $d_v'/d_v$, complexity depends on the private code parity-check matrix
density.

\ac{LDPC} decoding is performed through iterative algorithms working on the code Tanner graph, which has 
a number of edges equal to the number of ones in the code parity-check matrix.
Hence, for a given $d_v'$, the choice of $m$ and $d_v$ represents a tradeoff between complexity
of the steps i) and ii): increasing $d_v$ (and decreasing $m$, at most down to $1$, as in \ac{MDPC} code-based variants)
decreases the complexity of the step i) and increases that of the step ii), while increasing $m$ (and decreasing $d_v$,
as in \cite{Baldi2008, Baldi2012}) increases the complexity of the step i) and decreases that of the step ii).

In order to assess this tradeoff, we define two compact complexity metrics for steps i) and ii): $n m$ is
the number of operations needed to perform multiplication of a vector by $\mathbf{Q}$ and $n d_v I$, where
$I$ is the average number of decoding iterations, is proportional to the number of operations needed to
perform \ac{LDPC} decoding. In order to provide the actual count of binary operations, the latter should
be further multiplied by the number of binary operations ($\alpha$) performed along each edge of the Tanner graph.
However, this quantity depends on the specific decoding algorithm used.
In order to keep our analysis as general as possible, we first consider $\alpha=1$, and we will comment on the
effect of higher values of $\alpha$ later on.

Since $d_v = d_v' / m$, optimizing the tradeoff between steps i) and ii) reduces to choosing $m$ which 
minimizes:
\begin{equation}
C(m) = n \frac{d_v'}{m} I + nm.
\label{eq:C}
\end{equation}
This must be performed by considering a value of $d_v'$ able to guarantee sufficient security
against \ac{DCA} (see Fig. \ref{fig:DCA}) and a value of $n$ such that the code is able to correct $t' = m t$ errors, where $t$
is chosen in such a way as to reach a sufficient security level against \ac{ISDA} (see Fig. \ref{fig:ISDA}).

We observe that the minimum of \eqref{eq:C} corresponds to $m' = \sqrt{d'_v I}$.
However, for $m = m'$, the private code might be unable to correct all $m t$ errors, hence a smaller
value of $m$ might be necessary.
In addition, a high value of $m$ implies a small $d_v$ and, if $d_v$ becomes too small, the
private parity-check matrix could be discovered by enumeration.
On the other hand, by decreasing $m$ below $m'$, the value of \eqref{eq:C} increases, and reaches
a maximum for $m=1$, which is the minimum $m$ allowed to have a non singular matrix $\mathbf{Q}$.
Based on these considerations, we can conclude that the optimum value of $m$ is always greater than $1$,
and comprised between $1$ and $m'$. By considering a more sensible value of $\alpha > 1$, $m'$ would further increase.
However, this would have no effect on the actual optimal value of $m$, which, for the system parameters
that are of practical interest, always remains below $\sqrt{d'_v I}$.

%The main question is whether it is convenient to use $m=1$ (and $d_v' = d_v$), as in \ac{MDPC} code-based proposals \cite{Misoczki2012, Biasi2012},
%or rather to increase $m$ (and decrease $d_v$), as in previous \ac{LDPC} code-based proposals \cite{Baldi2008, Baldi2012}. 
%In order to have some indication on this, we can consider the ratio:
%\begin{equation}
%\frac{C(1)}{C(m)} = m \frac{d_v' I + 1}{d_v' I + m^2}.
%\label{eq:Cratio}
%\end{equation}
%We observe that, for $m \rightarrow \infty$, the ratio in \eqref{eq:Cratio} tends to zero,
%hence fixing very large values of $m$ is not advantageous.
%On the other hand, by fixing moderate values of $m > 1$, it may happen that $d_v' I \gg m^2$,
%in which case the ratio in \eqref{eq:Cratio} is approximately equal to $m$.
%Hence, this produces a complexity reduction on the order of $m$ with respect to the choice $m=1$.
%Some practical examples in this sense are provided in the next section.

Finally, we also observe that a low value of $m$ also affects the
total number of different matrices which can be chosen as $\mathbf{Q}$.
When $m=1$, the matrix $\mathbf{Q}$ becomes a \ac{QC} permutation matrix $\mathbf{P}$, that is, a matrix
formed by $n_0 \times n_0$ circulant blocks with size $p$, among which only one block per row
and per column is a circulant permutation matrix, while all the other blocks are null.
Hence, the total number of different choices for $\mathbf{P}$ is $p^{n_0} n_0!$.
For example, by considering the parameters proposed in \cite{Misoczki2012} for achieving
$80$-bit security, which are $(p=4800, n_0=2)$, $(p=3584, n_0=3)$ and $(p=3072, n_0=4)$, we would
have, respectively, $2^{25.46}$, $2^{38.01}$ and $2^{37.34}$ different choices for $\mathbf{P}$,
which would be too few to guarantee security.
However, this weakness can be avoided by resorting to a CCA2 secure conversion of the system,
and hence eliminating $\mathbf{S}$ and $\mathbf{P}$, as pointed out in \cite{Misoczki2012}.
On the other hand, when using higher values of $m$, this potential weakness can easily be avoided, just
for moderately high values of $n_0$ (like $n_0 = 3,4$), as needed for achieving
high code rates.

\section{Design examples}
\label{sec:Examples}

%In order to provide some design examples, we first refer to the parameters suggested in \cite{Misoczki2012}
%for achieving $128$-bit security with $n_0=4$, which are $p=6272$ ($n=25088$), $d_v' = d_v = 85$ ($m=1$), and $t=68$.
%According to Fig. \ref{fig:BFthresholds}, and considering that the \ac{BF} threshold underestimates the actual
%error correction capability, a code with the same parameters and $d_v = 29$ (that is, $m \approx 2.9$) can be
%actually able to correct $t' = m t = 197$ errors.
%Hence, the two solutions achieve the same security level by using different densities of the private parity-check
%matrix. In order to compare the complexity, we refer to \eqref{eq:C}.
%Typical values of the average number of decoding iterations are about $10$ \cite{Baldi2012}; so, by using $I=10$
%in \eqref{eq:C}, for these two cases we obtain $C(1) = 2^{24.35}$ and $C(2.9) = 2^{22.82}$.
%Hence, using a higher value of $m$ results in a complexity reduction of about three times.

We first consider the target of $100$-bit security.
According to Figs. \ref{fig:DCA} and \ref{fig:ISDA} (and assuming the shortest code length there considered, which provides
a conservative estimate), this can be achieved, with $n_0=4$, by choosing $d_v' = 59$ and $t = 47$.
An \ac{MDPC} code with length $n=16384$ and $d_v = 59$ has a \ac{BF} threshold equal to $68$ errors,
and we have verified that it is actually able to correct $47$ errors with very high probability.
Hence these parameters provide a $100$-bit security system design with $m=1$.
Instead, if we fix $d_v = 15$ (that is, $m=3.93$), we have $t' = 185$.
From Fig. \ref{fig:BFthresholds} it results that an \ac{LDPC} code with $d_v = 15$ and $n=16384$ 
has a \ac{BF} threshold equal to $187$ errors, and we have shown in Section \ref{sec:ErrCorr}
that, for such sparse codes, the \ac{BF} threshold actually provides a conservative estimate
of the number of correctable errors.
So, we have two system designs which achieve the same security level, but with different
matrix densities.
In these two cases, and by considering that a typical value of $I$ is $10$,
we have $C(1) = 2^{23.21}$  and $C(3.93) = 2^{21.27}$.

As another example, we consider a $128$-bit security level.
Similarly to the previous case, from Figs. \ref{fig:DCA} and \ref{fig:ISDA} we obtain that this
requires $d'_v=77$ and $t=62$.
An \ac{MDPC} code-based design can be obtained with code length $n=28672$ (and $d_v = 77$),
which provides a \ac{BF} threshold equal to $98$ errors.
We have verified that such an \ac{MDPC} code is actually able to correct $62$ errors with very
high probability, hence this solution reaches $128$-bit security with $m=1$.
An \ac{LDPC} code-based alternative can be obtained by using the same code length and
$d_v = 15$, that is, $m=5.13$. In this case, the \ac{BF} threshold is equal to $327$ errors,
hence the code is able to correct all the $t' = 318$ errors with very high probability.
In these cases (and with $I=10$), we have $C(1) = 2^{24.40}$ and $C(5.13) = 2^{22.09}$.

These examples confirm that, for a fixed security level, choosing sparser codes, and hence
higher values of $m$, is advantageous from the complexity viewpoint.

\section{Conclusion}
\label{sec:Conclusion}

In this paper, we have analyzed the choice of the private parity-check matrix density in
\ac{QC-LDPC} code-based variants of the McEliece cryptosystem.
We have shown that a given security level can be achieved by a balancing of the density
of the private parity-check matrix $\mathbf{H}$ and that of the matrix $\mathbf{Q}$ used to
disguise $\mathbf{H}$ into the public key.

Through some practical examples, we have shown that, from the complexity standpoint, it is 
generally preferable to decrease the density of the private parity-check matrix and to increase
that of the transformation matrix $\mathbf{Q}$.
For this reason, \ac{LDPC} code-based instances of the system result to be preferable to
\ac{MDPC} code-based instances if one wishes to keep complexity at its minimum, for a fixed
security level.

\newcommand{\BIBdecl}{\setlength{\itemsep}{0.01\baselineskip}}
\bibliographystyle{IEEEtran}
\bibliography{Archive}

\end{document}